\documentstyle[epsf]{mn}

\newcommand{\doverd}[2]{\frac{\partial #1}{\partial #2}}

\newcommand{\subscr}[1]{_{\rm #1}}
\newcommand{\superscr}[1]{^{\rm #1}}

\begin{document}

\title{On relativistic discs and rings}
\author[W. Kley]
       {W. Kley\\
        Max-Planck-Society, Research Unit Gravitational Theory,
        Universit\"at Jena, Max-Wien-Platz 1, D-07743 Jena, Germany\\
         }

\maketitle

\begin{abstract}
Sequences of infinitesimally thin,
uniformly rotating, self-gravitating  relativistic discs with
internal two-dimensional
pressure have been constructed. It is shown that in weaker relativistic
configurations the sequences undergo a continuous bifurcation from a disc
to a ring structure,
while in stronger relativistic cases the sequences terminate
at the mass-shed limit where gravitational forces are exactly balanced
by centrifugal forces.
\end{abstract}


\section{Introduction}
Relativistic, self-gravitating discs may have played an important
role during the process of galaxy formation in the early universe. 
For example, they can
be considered as possible predecessors of galaxies today that
contain a massive black hole in their centre.
Also, during the final evolution of a close binary system, consisting
of two neutron stars, the orbit shrinks rapidly due to the
emission of gravitational waves and the two stars merge finally
into one object.
Shortly after this coalescence, the system can be approximately described
as a highly flattened disc-like structure which may display relativistic
effects.
These systems may then be modeled, in a simplified approach, by
infinitesimally thin discs where the vertical variation has been averaged
onto the midplane.  

Uniformly rotating relativistic discs with zero internal pressure
have been studied in detail by Bardeen \& Wagoner (1971). Here
an expansion method was used in which the problem was reduced to a set of
ordinary differential equations, which were then solved numerically for the
higher order corrections. Nevertheless, using this method, strongly
relativistic disc structures could also be calculated accurately. 
Recently, the analytical solution of this problem in terms
of integral equations was obtained (Neugebauer \& Meinel 1993), which could
subsequently be solved utilizing hyper-elliptic functions
(Neugebauer \& Meinel 1995). This solution represents the first analytic
solution of a continuously extended rotating object in general relativity. 
In these approaches a given equation of state ($p=0$) and
rotation law ($\Omega=$const.) are assumed, and
the Einstein-equations are solved directly.

The method of mirror images is another frequently used option to
construct models of relativistic discs.
Here one begins with a known axisymmetric
metric (e.g. Kerr), makes a cut in the upper half of the $\rho-z$ plane
where $\rho$ and $z$ are cylindrical coordinates, 
and identifies the metric functions there with those of a symmetric cut
in the lower half which joins the first at the $z=0$ plane
(see Pichon \& Lynden-Bell 1996, and references therein).
This method yields disc solutions where the corresponding matter properties,
i.e. the energy-momentum surface distribution,
must be inferred from the pasted solution after the two cuts
have been identified.
The method was first used for relativistic discs by Turakulov (1990).

It is known that, in the Newtonian limit, the uniformly rotating
pressure-less dust disc is
unstable to all perturbing modes (Binney \& Tremaine 1987),
a result which most likely carries over into the relativistic
regime (Bardeen \& Wagoner 1971). The addition of pressure tends to
stabilize the disc. For Newtonian discs it was shown recently (Kley 1996)
that uniformly rotating, pressurized  discs
undergo a continuous bifurcation process into a ring-like structure via
a sequence of dumb-bell shaped equilibria.
This process has its three-dimensional analogue in the bifurcation
of rotating stars into rings (Eriguchi \& Sugimoto 1981).

In this paper we generalize the relativistic zero-pressure disc solutions by
imposing a given (polytropic) equation of state relating the surface pressure
with the surface density.
This pressure may be thought of as being generated by the
random internal peculiar velocities in a galaxy, and it has
a stabilizing effect.
This work on the other hand also generalizes the Newtonian results
on thin Maclaurin discs and their bifurcations to the relativistic case.

In Section 2, the relevant equations are given and the numerical method is
outlined in Section 3. In Section 4 the results of the numerical computations
are presented and in Section 5 we conclude.

\section{Basic equations}
\subsection{Einstein equations}
We shall treat the problem of an infinitesimally thin gaseous disc with
internal pressure acting only in the plane of the disc. The problem is
treated in full general relativity.
The disc is assumed to rotate rigidly with the angular velocity $\Omega$.
We use a cylindrical coordinate system
\begin{equation}
  x_1=\rho, \, x_2=z, \, x_3=\varphi, \, x_4=t
\end{equation}
where the disc is located in the equatorial plane ($z=0$).
The line element is written in the
form used by Bardeen \& Wagoner (1971)
\begin{equation}
  ds^2 = e^{2 \mu} \left( d \rho^2 + dz^2 \right)
     + \rho^2 B^2 e^{-2 \nu} ( d \varphi - \omega dt)^2 - e^{2 \nu} dt^2.
\end{equation}
This metric describes general axisymmetric and stationary
configurations in general relativity.
Four metric functions ($\mu, \nu, B,$ and $\omega$),
which depend only on the coordinates $\rho$ and $z$, have to be determined.

For the energy-momentum tensor we use the expression for an ideal gas. 
We assume relativistic units and set $G=c=1$ in all formulae.
\begin{equation}  \label{Tab}
   T_{ab} = ( \epsilon + p ) u_a u_b 
      + p \left( g_{ab} - \frac{N_a N_b}{N^c N_c} \right),
\end{equation}
with
\begin{equation}
    N_a = \doverd{z}{x^a} = (0,1,0,0).
\end{equation}
In equation (\ref{Tab}), $\epsilon$ is the total mass-energy density
and $p$ is the gas pressure. These are three-dimensional quantities.
The last term, containing a projection along the $z$-axis, has
been introduced to ensure that the pressure acts
only in the equatorial plane, i.e. the forces perpendicular to
the disc (in the $z$-equation) vanish.

The four velocity in case of pure rotation (no radial or vertical motions)
is given by
\[  u^a = u^4 (0,0,\Omega, 1) \]
where $\Omega$ is the angular velocity as seen by an observer
at infinity. Throughout this work we assume rigid rotation, and thus
$\Omega$ is a constant.
From the normalization of the four velocity ($u_a u^a = -1$) one obtains
with the given metric
\begin{equation} 
    u^4 =\frac{e^{-\nu}}{\left( 1 - v_{\varphi}^2 \right)^{1/2}},
\end{equation}
where
\begin{equation}
     v_\varphi = \rho B e^{-2 \nu} (\Omega - \omega)
\end{equation}
is the rotational velocity in the locally non-rotating frame of reference.

The Einstein equations $ G_{ab} = 4 \pi T_{ab}$ yield four
second order partial differential equations for the four metric
potentials (see also Bardeen \& Wagoner 1971)
\begin{eqnarray}  \label{v3d}
   e^{-2\mu} \left[ \frac{1}{B} \nabla(B \nabla \nu)  -
     \frac{1}{2} e^{-4\nu} \rho^2 B^2 \nabla \omega \cdot \nabla \nu \right]
    =  \nonumber \\
      4 \pi \left[ ( \epsilon + p) \frac{1+v_\varphi^2}{1-v_\varphi^2}
          + p \right],
\end{eqnarray}
\begin{eqnarray}  \label{w3d}
   \frac{e^{-2(\nu +\mu)}}{2 \rho B^2} \left[
         \nabla(\rho^2 B^3 \nabla \omega)
       - 4  \rho^2 B^3 \nabla \nu \cdot \nabla \omega  \right]
            = \nonumber \\
         - 8 \pi ( \epsilon + p) v_{\varphi}
        \left(1 - v_{\varphi}^2 \right)^{-1},
\end{eqnarray}
\begin{equation}  \label{B3d}
    \frac{e^{-2 \mu}}{\rho B} \nabla(\rho\nabla B)  =  8 \pi p, 
\end{equation}
and
\begin{eqnarray}  \label{m3d}
   \frac{e^{-2 \mu}}{2 \rho B} \left[- 4 \rho B
       ( \mu,_{\rho \rho} + \mu,_{z z} )
      + 4 \nabla e^\nu \cdot \nabla (B \rho e^{-\nu})
      + \right. \nonumber \\
  \left.  e^{-4 \nu} \rho^3 B^3 \nabla \omega \cdot \nabla \omega \right]
         = 8 \pi \epsilon.
\end{eqnarray}
Here the $\nabla$-operator is used in cylindrical coordinates ($\rho, z$).

Since the disc is located in the $z=0$ plane, it can be treated
mathematically as a boundary condition to the otherwise empty space.
The above equations reduce in the case of a vacuum to 
\begin{eqnarray} 
   \nabla(\nabla \nu) & = & \frac{1}{2} \rho^2 B^2 e^{-4 \nu}
          \nabla \omega \cdot \nabla \omega
        - \frac{1}{B} \nabla B \cdot \nabla \nu  \label{vvac}\\
   \nabla(\rho^2 \nabla \omega) & = & - \frac{\rho^2}{a}
          \nabla a \cdot \nabla \omega,  \label{wvac}\\
   \nabla(\rho\nabla B) & = & 0  \label{bvac}\\
   \Delta\subscr{c} \mu  & = & \frac{1}{4} \rho^2 B^2 e^{-4 \nu}
          \nabla \omega \cdot \nabla \omega
        + \frac{1}{B} \nabla B \cdot \nabla \nu \nonumber \\
     & - & \nabla {\nu} \cdot \nabla \nu  + \frac{1}{\rho} \doverd{\nu}{\rho}
           \label{mvac}
\end{eqnarray}
where $\Delta\subscr{c}$ denotes the Cartesian Laplace operator (for the
coordinates $\rho, z$), and
\[  a:= \frac{B^3}{e^{4\nu}}. \]

In the disc we define the two-dimensional total surface energy density
$\sigma$, and the surface pressure $P$ through
\begin{equation}
   \epsilon =\sigma(\rho) \delta(z) e^{- 2 \mu}, \\
   p = P(\rho) \delta(z) e^{- 2\mu},
\end{equation}
where $\delta(z)$ denotes the usual Dirac delta function.

The proper density and pressure in the disc are then given by
\begin{equation} \label{proper}
   \sigma\subscr{p} = \sigma e^{-\mu}, \\ 
   P\subscr{p} = P e^{-\mu}.
\end{equation}

From momentum and energy conservation ${T^{ab}}_{;b}=0$ we obtain
the radial hydrostatic equation in the plane of the disc
\begin{equation}  \label{hydro1}
  P\subscr{p,\rho} + (\sigma\subscr{p} + P\subscr{p}) U'_{,\rho}  = 0,
\end{equation}
where $U'$ is defined through
\[ e^{2 U'} = \left(u^4\right)^2  
    = \left(1 - v_\varphi^2\right)^{-1} e^{-2 \nu}.  \]
In deriving (\ref{hydro1}) we assumed a constant rotation rate $\Omega$.

The total angular momentum of the disc is given by
\begin{equation}  \label{J}
    J = 2 \pi \int\subscr{Disc} ( \sigma + P )
      \frac{v_\varphi}{1 - v_\varphi^2} e^{-2 \nu} B^2 \rho^2 {\rm d} \rho
\end{equation}
and the total gravitational mass by
\begin{equation} \label{M}
    M = 2 \pi \int\subscr{Disc} ( \sigma + 2 P )
      B \rho {\rm d} \rho  +  2 \Omega J
\end{equation}
\subsection{Equation of state}
To close the set of equations, thermodynamic relations have to be added.
We present calculations for two different equations of state. In the first,
more explorative, test case we assume a simple relation between the
total proper surface energy density $\sigma\subscr{p}$ and
the proper (surface) pressure $P\subscr{p}$ 
\begin{equation}  \label{eos0}
     P\subscr{p} = \sigma\subscr{p}^{\gamma}, 
\end{equation} 
where $\gamma$ denotes a polytropic exponent which is set to 3 for all models.
This first equation of state has been used only for the non-rotating
models.
 
For the non-rotating and all rotating models we assumed an
equation of state that obeys, in analogy
to three-dimensional configurations, an isentropic relation
for the proper two-dimensional surface quantities
\begin{equation}
     {\rm d} \sigma\subscr{p} - \frac{\sigma\subscr{p} 
   + P\subscr{p}}{\sigma^0\subscr{p}} \, {\rm d} \sigma^0\subscr{p} = 0,
\end{equation} 
where $\sigma^0\subscr{p}$ denotes the proper surface rest mass density. The
the total surface density $\sigma\subscr{p}$ is given by
\[  \sigma\subscr{p} = \sigma^0\subscr{p} + \sigma\superscr{i}\subscr{p},\]
where $\sigma\superscr{i}\subscr{p}$ denotes the
proper (surface) internal energy density.  
In both cases equation (\ref{hydro1})
can be integrated and yields the condition of hydrostatic equilibrium.
For the isentropic case one obtains
\begin{equation} \label{hydstat}
   e^{\nu} \sqrt{ 1 - v_{\varphi}^2 }
    \, \frac{\sigma\subscr{p} + P\subscr{p}}{\sigma^0\subscr{p}}  = const.
\end{equation}
For the isentropic equation of state we assume further a polytropic relation
\begin{equation}  \label{eos1}
     P\subscr{p} = K \left(\sigma^0\subscr{p}\right)^\gamma 
         = (\gamma - 1) \sigma\superscr{i}\subscr{p}. 
\end{equation}

In the non-relativistic limit the metric potential $\nu$ turns into the
Newtonian potential and  
the hydrostatic equation reduces to the well known
relation
\[   \nu  + \frac{\gamma}{\gamma-1} \frac{P}{\sigma}
    - \frac{1}{2} r^2 \Omega^2  =  const.
\]
For $\gamma=3$ there exists an analytic solution in the Newtonian limit,
the Maclaurin disc solution (cf. Binney \& Tremaine 1987).
In this case the rotation rate $\Omega$ and the polytropic constant
$K$ are related through
\begin{equation} \label{newtok} 
      \frac{K}{K\subscr{max}} = 
   1 - \left( \frac{\Omega}{\Omega\subscr{c}} \right)^2,
\end{equation}
where $K\subscr{max}$ is the maximum value of $K$ which is taken for
zero rotation $\Omega=0$, and $\Omega\subscr{c}$ is the maximum rotation
rate for the pressure-less dust disc which has $K=0$.
The surface density is given by
\begin{equation} \label{newtsig}
    \sigma(\rho) = \sigma_0 \sqrt{ 1 - \left(\rho/\rho\subscr{d}\right)^2},
\end{equation}
where $\sigma_0$ denotes the central surface density of the disc and
$\rho\subscr{d}$ is the disc radius.
\subsection{Boundary conditions}
In the equatorial plane ($z=0$) the potentials are continuous.
The presence of the disc however is given by jump conditions for the
first derivatives of the metric potentials. By vertically
integrating equations (\ref{v3d} to \ref{m3d}) the following
jump conditions are obtained:
\begin{eqnarray}
   \nu_{,z} & = & 2 \pi \left[ (\sigma + P) 
      \frac{1 + v_{\varphi}^2}{1 - v_{\varphi}^2}  + P \right]  \label{vjmp}\\
   \omega_{,z} & = & - 8 \pi (\sigma + P) 
      \frac{\Omega - \omega}{1 - v_{\varphi}^2}  \label{wjmp}\\
   \frac{B_{,z}}{B} & = & 4 \pi P   \label{bjmp}\\
   \mu_{,z} & = & -  2 \pi \sigma \label{mjmp}.
\end{eqnarray}
These equations form the boundary conditions within the disc region 
$\rho < \rho\subscr{d}$.
Outside, for $\rho > \rho\subscr{d}$ all vertical derivatives with
respect to $z$ vanish. On the polar axis ($\rho = 0, z > 0$) the
radial derivatives of all potentials are zero. Because of local
flatness the additional relation
\begin{equation}
    B = e^{\nu + \mu}
\end{equation}
holds on the axis, and will be used in obtaining the numerical solution. 
At infinity the potentials $\nu, \omega, \mu$ vanish,
and $B=1$. 

\section{Method of solution}
The vacuum Einstein equations (\ref{vvac}-\ref{mvac}), together with
the boundary conditions (\ref{vjmp} - \ref{mjmp}) in the disc, form a very
complicated set of non-linear partial differential equations and we resort
to a numerical solution.
As it is numerically inconvenient to work with boundary conditions
at infinity, we first perform a coordinate transformation that takes infinity
to a finite distance, say 1.
Specifically, we transform the equations into the new coordinates
$u$ and $v$ which are defined by:
\begin{equation}
     u = \frac{\rho}{\rho+\rho\subscr{d}}, 
     \hspace{0.5cm} \mbox{and} \hspace{0.5cm}
     v = \frac{z}{z+\rho\subscr{d}}.
\end{equation}
The disc has the radius $u\subscr{d}=0.5$ in these coordinates, which
is held fixed in all models.

To obtain the actual numerical solution, the partial
differential equations (\ref{vvac}-\ref{mvac}) are discretized on
a suitable grid of $NU \times NV$ grid-cells.
In the models presented here, a resolution of
$128 \times 128$ grid-cells was used; only in the stronger relativistic cases
was a higher resolution of $256 \times 256$ used.
The radial grid-points are usually spaced equidistantly in physical
space within the disc region $\rho < \rho\subscr{d}$ ($u < 0.5$) and
equidistantly in $u$ outside this region. The vertical grid-points are spaced
equidistantly in $v$.
Specifically, the first radial grid-point
$u_1$ lies at $u=0~(\rho=0)$ and the last one $u_{NU+1}$ at
$u=1~(\rho=\infty)$, and similarly for the $v$-coordinate.
The discretization leads to a set of four coupled systems of linear
equations which have a banded structure. The non-linear right hand sides
of eqs. (\ref{vvac}-\ref{mvac}) are not linearised, but rather updated
at each cycle during the process of iteratively solving
the matrix-equations.
This iterative solution of the matrix equations is obtained by the
successive over relaxation (SOR) method,
which is easy to program and allows a simple treatment
of the boundary conditions.

The boundary conditions at infinity and on the axis are implemented
straightforwardly. At infinity the values of $\nu, \omega$, and $\mu$ are
set to zero and $B$ is set to 1. On the
axis and outside the disc the derivatives vanish which is
implemented, for example, to second order in the grid spacing as
\[     \nu_{1j} = ( 3 \nu_{3j} - \nu_{2j} ) / 4, \]
where the index $j$ denotes the vertical grid-points.
Only the actual jump conditions within the disc
($v=0, u \le u\subscr{d}$) have to be treated with more care.
Starting from the minimum radius of the disc (ring) and then
moving subsequently
to larger values, the updated surface density is calculated from the
hydrostatic equation (\ref{hydstat})
using the old values of the metric potentials.
The jump conditions (\ref{vjmp} - \ref{mjmp}) are then used to obtain
the new values of the potentials in the disc.
Using this procedure, the boundary values are updated at each
iteration cycle of the solver of the matrix equation.

The procedure of the solution depends on the type of solution sought.
For the cases of disc and ring, and also for the intermediate
`dumb-bell' case which connects these two,
slightly different methods have to be applied (see Kley 1996).
In all cases the outer radius $\rho\subscr{d}$ of the configuration
is held fixed at the value $u\subscr{d} = 0.5$ in dimensionless units.
Additionally, the central potential $\nu\subscr{c} = \nu(0,0)$ is
fixed during the iterations which is the equivalent of fixing the central
redshift $z$, which is given by
\begin{equation}
	z = \frac{1}{\sqrt{-g_{tt}(0,0)}} - 1.
\end{equation}
where
\begin{equation} 
     g_{tt}= \left(\rho B e^{-\nu}\omega\right)^2 - e^{2\nu}.
\end{equation} 
\epsfverbosetrue
\begin{figure}
\epsfxsize=8.5cm
\epsfbox{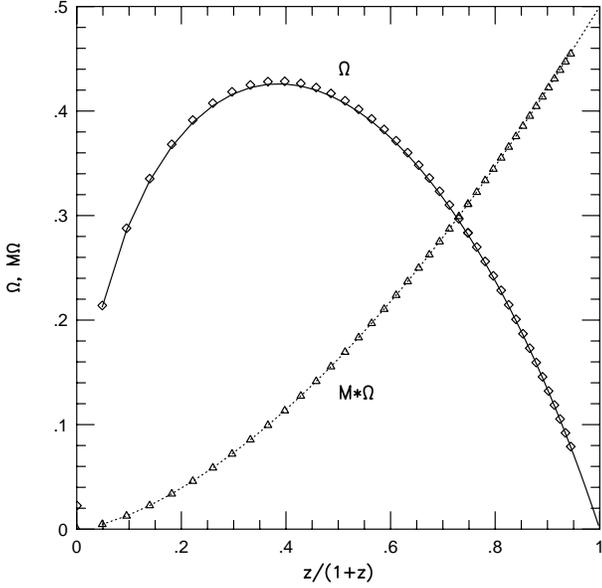}
  \caption{The angular velocity $\Omega$ and the total gravitational
  mass $M$ as function of central redshift $z$ for the pressure-less
  dust disc. The symbols refer to the numerical solution and the lines
  to the analytical solution.}
 \end{figure}
\epsfverbosetrue
\begin{figure}
\epsfxsize=8.5cm
\epsfbox{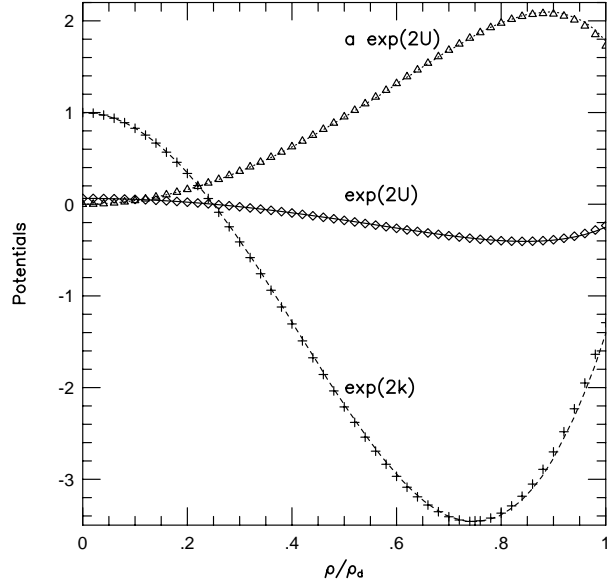}
  \caption{Comparison of numerical and analytical results
  for three metric functions (see text) for the dust disc having
  a central redshift of $z=2.97$.
  Lines refer to the analytical solution and symbols to the numerical
  solution. The radial coordinate $\rho$ is given in units of the
  disc radius $\rho\subscr{d}$.}
 \end{figure}
The problem is then to determine the rotational angular velocity
$\Omega$ and the pressure constant $K$ simultaneously during the iterations.
In the disc regime, either $\Omega$ or $K$ can be fixed, and the other
parameter can then be calculated from the hydrostatic equation (\ref{hydstat}).
In the case of fixed $\Omega$, the constant in (\ref{hydstat}) is fixed at the
disc radius $\rho\subscr{d}$ and $K$ is then obtained from evaluating
(\ref{hydstat}) at the centre of the disc.
In regions where $K(\Omega)$ may be double valued, which occurs in fact
in the vicinity of the bifurcation to a dumb-bell sequence, it may
be advantageous to switch the procedure and fix $K$. In this case $\Omega$
is calculated from (\ref{hydstat}) at a point within the disc, as
at the origin the rotational velocity is zero. In the disc region both methods
lead to the same results, which provides a good test for the internal
consistency of the
method. The latter method of holding $K$ fixed during the iterations can in
fact used up to that point at which the central density of the disc
has reached zero.
This dumb-bell region can also be calculated by fixing the central density
$\sigma(0)$ and obtaining $\Omega$ and $K$ simultaneously from the
hydrostatic equation (\ref{hydstat}) at two different locations.
Again, this is consistent with the previous method.
The dumb-bell part ends with a model which has zero central density
at $\rho=0$, and from here on the ring solutions continue.
In this sequence, in addition to the
outer radius $\rho\subscr{d}$, the inner radius $\rho\subscr{i}$ of the ring
is also held fixed. As before, $K$ and $\Omega$ are obtained
from the hydrostatics.
The global accuracy of the method can be tested by comparing the
asymptotic behaviour of $\nu$ and $\omega$ with
the total mass and angular momentum of the disc.
As the spherical radial coordinate 
$r = (\rho^2 + z^2)^{1/2} \rightarrow \infty$
\begin{eqnarray}
      \nu & \rightarrow & - \frac{M}{r} \\
      \omega & \rightarrow & - 2 \frac{J}{r^3}.
\end{eqnarray}
Numerically, the evaluation of $M$ and $J$ at large distances agrees
with the values obtained by integration over the disc (\ref{J}) and
(\ref{M}), to within one percent. Only for stronger relativistic discs the
agreement becomes weaker, due to the strong mass concentration in the
centre of the disc. Additional test cases for the numerical method
are presented below in the results section.
\section{Results}
\subsection{Test case: The pressure-less disc}
Ideally, any newly developed numerical method needs to be
tested against known analytical
solutions. Fortunately, there exists a solution in the special case of
a fully relativistic pressure-free disc (Neugebauer \& Meinel 1993, 1995).
A sequence of dust disc models were run with a varying central
redshift. In Fig.~1 the angular velocity $\Omega$ (solid-line, diamonds)
and the global quantity $M \Omega$ (dotted-line, triangles) are compared
with the analytical solution for different central redshift $z$.
The numerical and analytical solutions agree very well. 
Note that the angular velocity $\Omega$ is not a given fixed input parameter,
but is determined self-consistently as a function of $z$ (with fixed
outer disc radius $\rho\subscr{d}$) during the iterative solution. 
The normalized central redshift $z\subscr{n}=z/(1+z)$ serves as parameter which
measures the strength of relativistic effects. In the Newtonian limit
$z\subscr{n}=0$ and in the extreme relativistic limit $z\subscr{n}=1$.
%
\epsfverbosetrue
\begin{figure}
\epsfxsize=8.5cm
\epsfbox{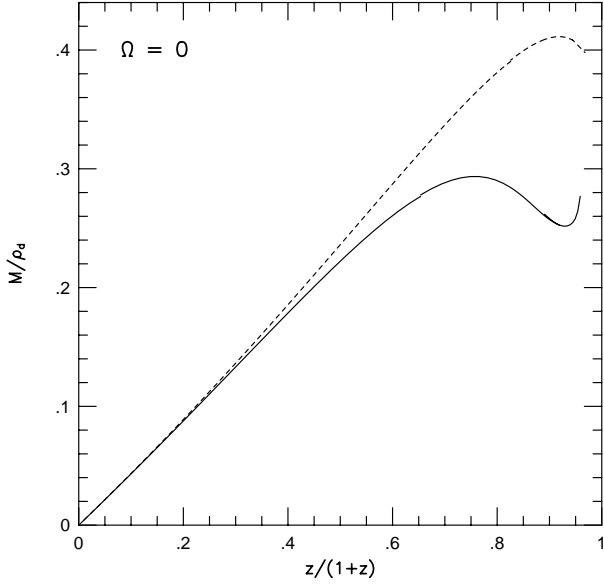}
  \caption{The total gravitational mass $M$ versus central redshift
  for the non-rotating disc. The dashed lines refers to the equation of state
  (20) and the solid line to (23).}
 \end{figure}
\epsfverbosetrue
\begin{figure}
\epsfxsize=8.5cm
\epsfbox{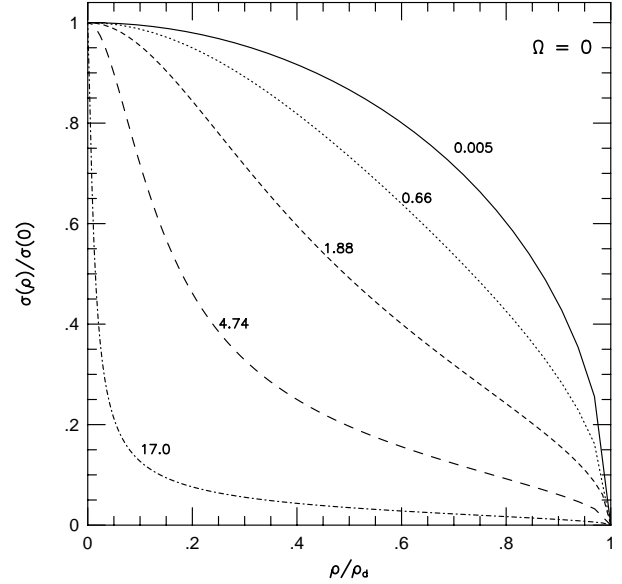}
  \caption{The normalized surface density $\sigma$ as a function of
  radius for the non-rotating disc. The labels denote the value of the
  central redshift of the disc.}
 \end{figure}
%
\epsfverbosetrue
\begin{figure}
\epsfxsize=8.5cm
\epsfbox{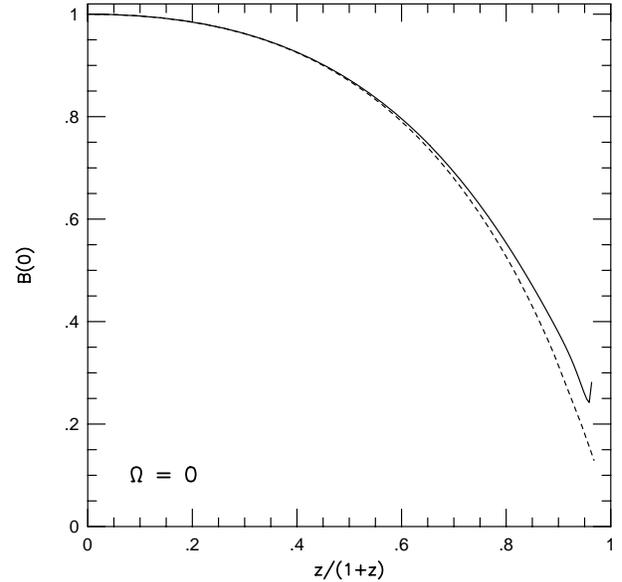}
  \caption{The central value of the potential $B$ versus central redshift
  for the non-rotating disc for two different equations of state
 (see fig.~3).}
\end{figure}
Locally, for the metric potentials, the agreement is very good as well.
The metric potentials displayed in Fig.~2 are those of
the line element written in Weyl-Lewis-Papapetrou form which has been
used by Neugebauer \& Meinel (1993). They are related to $\nu, \omega, \mu$ 
by
\begin{eqnarray}
      e^{2 U} & = & e^{2 \nu} - \rho^2 \omega^2 e^{-2 \nu} \\
      a e^{2 U} & = & \rho^2 \omega e^{-2 \nu} \\
      e^{2 k} & = & e^{ 2 (\nu + \mu)}.
\end{eqnarray} 
The potential $B$ is equal to 1 everywhere for the dust disc.
The results are compared for an intermediate redshift $z=2.97$.
For the general case, with internal pressure, there exists
no analytical solution for a relativistic disc. In the Newtonian limit
however, the numerically obtained solutions also agree very well with the
known analytic Maclaurin disc solution (\ref{newtok}, \ref{newtsig})
and with the bifurcation diagrams
found in the Newtonian case (Kley 1996). Although a complete test
of the numerical approach was not possible,
the test cases provide very positive indication as to the
reliability of the results.
\subsection{Non-rotating discs}
The non-rotating discs with $\Omega=0$ are supported purely by
internal pressure and have, for sequences of a given central
redshift $z$, the maximum value of $K$.
Similar to stars, non-rotating discs have a maximum mass 
$M/\rho\subscr{d}$ which increases with central redshift.
In Fig.~3 the total gravitational mass is plotted versus central
redshift for the two different equations of state.
The higher total masses are reached in the case of 
the simple equation (\ref{eos0}), dashed line, while the polytropic relation
(\ref{eos1}) has a maximum at a finite redshift $z \approx 3.25$. The stronger
relativistic cases have been calculated here using a higher grid resolution
of $256 \times 256$. However, the last part of the curves for
$z/(1+z) \ga 0.9$
may nevertheless still be unreliable. The problem arises from the strong
central concentration of the surface density (Fig.~4). Due to this problem
$M$ reaches its maximum for the first equation of state
(\ref{eos0}) most likely at infinite central redshift.
At larger and larger
redshifts, the density increasingly peakes at the origin
(refer to the model with $z=17.0$). The sequence
terminates eventually at a black hole configuration,
with all the mass located at the origin.
\epsfverbosetrue
\begin{figure}
\epsfxsize=8.5cm
\epsfbox{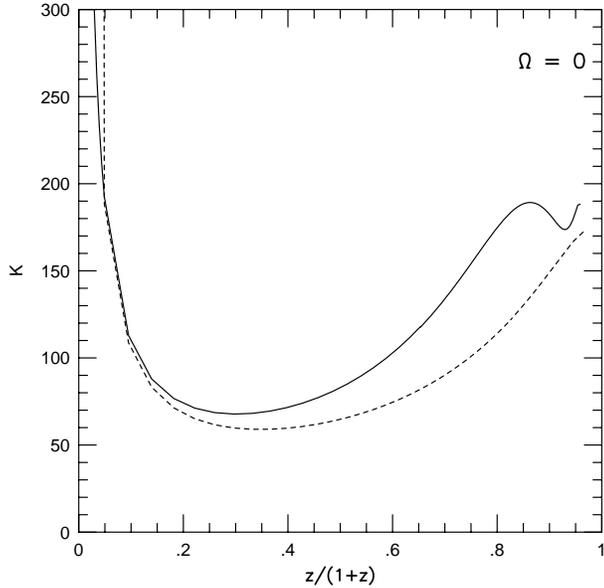}
  \caption{The polytropic constant $K$ versus central redshift for the
  non-rotating disc for two different equations of state (see Fig.~3).}
 \end{figure}
The strong increase of internal pressure with $z$ leads also to a
central drop in the metric potential $B$ which is plotted in Fig.~5. 
For the pressure-free dust disc, $B=1$ everywhere;
however, in the case of non-zero pressure,
equation (\ref{bjmp}) creates a deviation
from this simple relation. With increasing $z$, the metric potential
$B$ decreases (in all space) and its central value approaches zero in the
extreme relativistic limit ($z=1$).
The variation of the polytropic constant $K$ with redshift is displayed in
Fig.~6. With increasing redshift, $K$ drops from the Newtonian limit
and increases again for stronger relativistic discs. Adding
rotation ($\Omega>0$) will lower the value of $K$. In the following diagrams
$K$ will be normalized to this maximum value $K\subscr{max}$.
\subsection{Rotating discs}
%
\epsfverbosetrue
\begin{figure}
\epsfxsize=8.5cm
\epsfbox{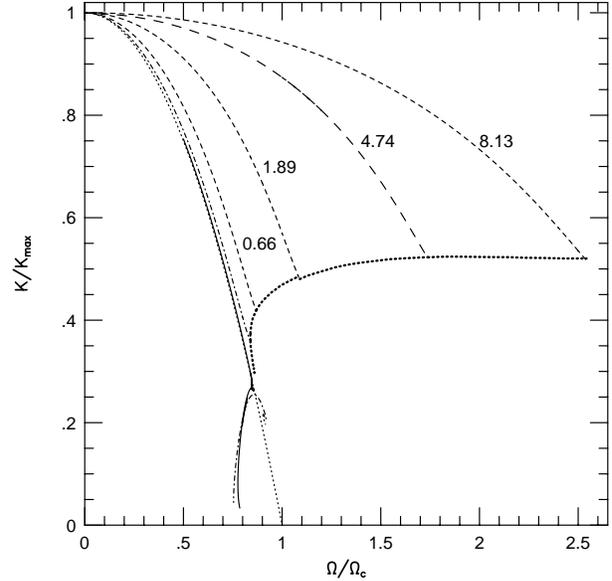}
  \caption{Pressure constant $K$ 
  versus angular velocity $\Omega$ given in units
  of the angular velocity of a dust disc $\Omega\subscr{c}$
  with the same central redshift. 
  The solid line refers to a central redshift $z=0.005$, and the
  dashed dotted line to $z=0.22$. The other curves are labeled with the
  corresponding $z$ values. The thin dotted line is the Newtonian limit
  (Maclaurin disc). The thick dotted line denotes the mass-shed limit.
  The values of $\Omega\subscr{c}$ and $K\subscr{max}$ can be read off from
  Fig.~1 and Fig.~6.}
 \end{figure}
%
\epsfverbosetrue
\begin{figure}
\epsfxsize=8.5cm
\epsfbox{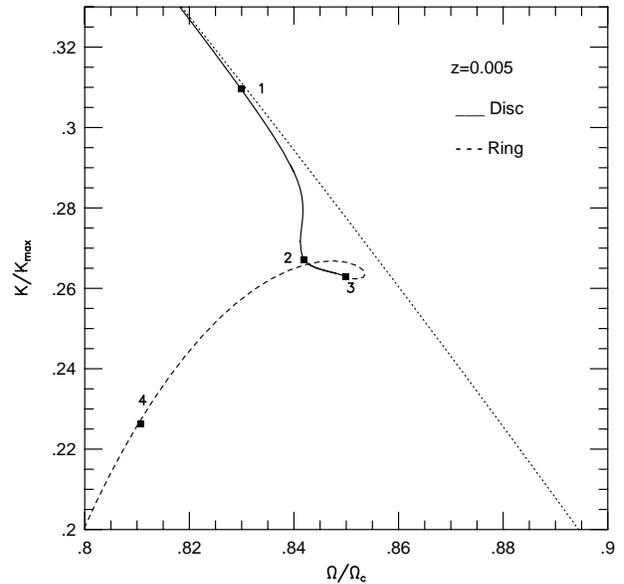}
  \caption{Enlargement of Fig.~7 displaying in detail the bifurcation
  from the disc to the ring configuration. The dotted
  line refers to the Newtonian disc solution. The solid line refers to
  the disc solution which turns into a ring structure (dashed line)
  at point 3. The corresponding surface density distributions
  at the marked locations are displayed in the next figure.}
\end{figure}
\epsfverbosetrue
\begin{figure}
\epsfxsize=8.5cm
\epsfbox{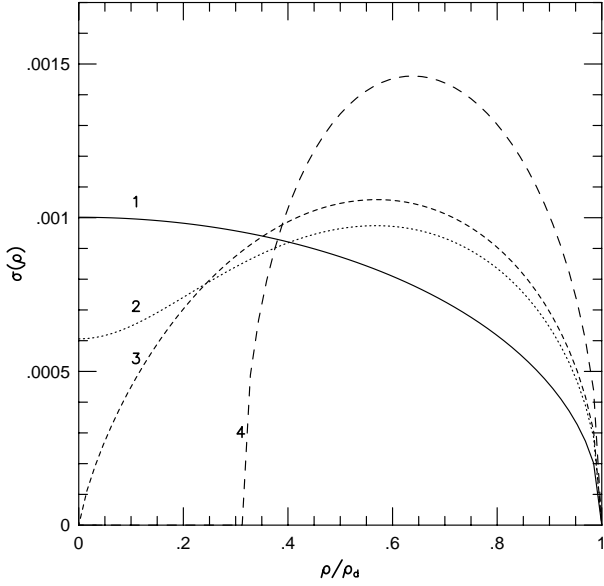}
  \caption{Plot of the surface density distribution versus radius
  at the locations marked in Fig.~7.
  Curve 1 follows closely the Newtonian curve}
\end{figure}
Rotating discs were studied for the isentropic second equation of state only
(\ref{eos1}), and the primary results are
displayed in Fig.~7 in the $K-\Omega$ diagram.
The curves are labeled with the corresponding central redshifts
of the sequences.
Weaker relativistic discs (solid line, $z=0.025$) follow
the Newtonian curve (thin dotted line, eq.~\ref{newtok})
closely up to a finite $\Omega < \Omega\subscr{c}$
and then bifurcate continuously
into a ring-like structure at $\Omega/\Omega_c \approx 0.84$.
This continuous transition from disc to ring occurs only for
$z \la 0.01$. The details of the bifurcation process will be explained
below.
For intermediate central redshifts $0.01 < z < 0.22$ (dashed-dotted line),
the discs and
rings coexist with no apparent connecting branch between them.
Stronger relativistic, $z \ga 0.22$, (dashed lines) discs terminate
in a mass shed limit, where at the outer edge of the disc
($\rho =\rho\subscr{d}$) gravity is balanced exactly by centrifugal forces.
From the hydrostatic equation we obtain the criterion for reaching
the mass-shed limit
\begin{equation} \label{mslimit}
    U'_{,\rho} 
      = \nu_{,\rho} - \frac{1}{2} \frac{v^2_{\varphi,\rho}}{1 - v^2_\varphi}
      = 0.
\end{equation}
The end points of the dashed curves and the thick dotted line
mark the loci where
the edge of the disc reaches this limit. In case of the dust disc, all radii
are at this limit simultaneously, since eq.~(\ref{mslimit})
is exactly the hydrostatic equation for a pressure-less disc.
It should be noted that the angular velocity reached
in these stronger relativistic
cases exceeds the value of the pressure-free case ($\Omega=\Omega_c$)
for the same central redshift. This effect is caused by the non-linearity
of the Einstein-equations. Through the gravitational action of the internal
pressure, the potential $B$ is then influenced, which in turn enters
through the rotational velocity $v_\varphi$ into the hydrostatic equation.

A more detailed view of the continuous
bifurcation process is given in Fig.~8, which
is an enlargement of the previous figure. 
The sequence bifurcates smoothly from the disc solution through a sequence
of dumb-bell shaped density distributions which show
a depression of the surface density in the middle region of the disc
(see Fig.~9). At point
3 the density in centre has reached zero, and the ring structures begin.
The numbers in Fig.~8 refer to the density distributions displayed
in Fig.~9. The solid line in Fig.~9 follows approximately the
Newtonian relation (\ref{newtsig}). With a growing inner
ring radius the maximum surface density increases, and the ring
sequence turns eventually into a line with zero radial extent.
These extreme ring models could not be obtained numerically, due to the lack
of resolution.
\epsfverbosetrue
\begin{figure}
\epsfxsize=8.5cm
\epsfbox{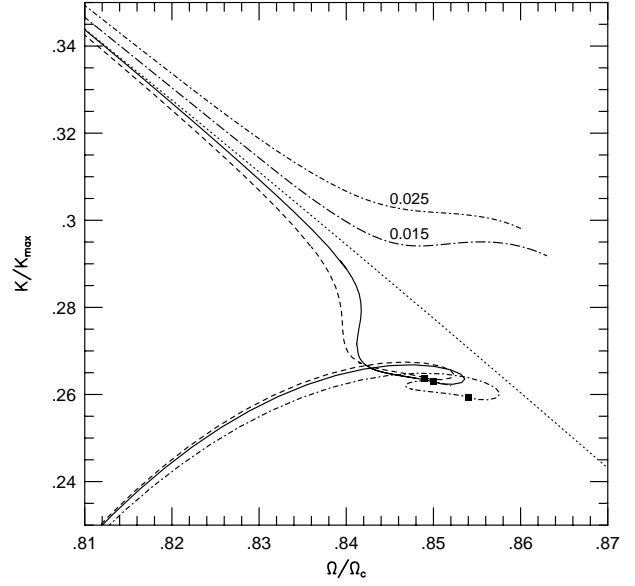}
  \caption{Detail of the bifurcation process for four different
  central redshifts (dashed line $z=0.0005$; solid line $z=0.005$, the
  others are given in plot). The Newtonian curve for the
  disc is given by the dotted line.
  The black squares mark the onset of the ring sequences (not plotted
  for $z=0.015$).} 
\end{figure}
\epsfverbosetrue
\begin{figure}
\epsfxsize=8.5cm
\epsfbox{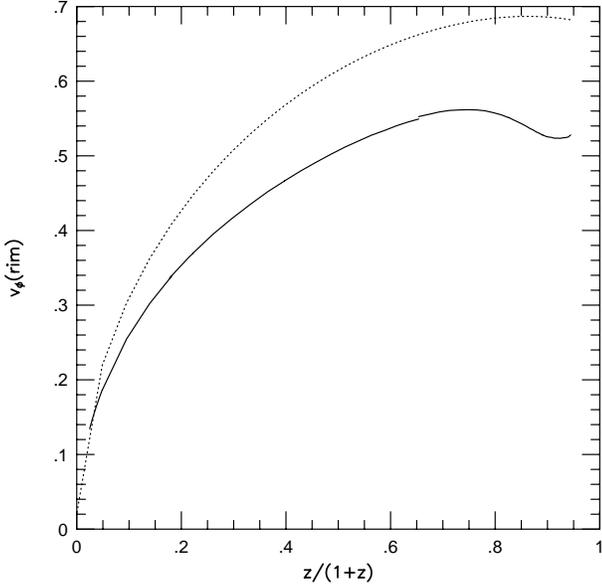}
  \caption{The rotational velocity at the rim ($\rho=\rho\subscr{d}$)
  of the disc as a function
  of central redshift for the dust disc (dashed line) and along
  the mass-shed limit (solid line). The curves join at the point where
  the disc begin to reach the mass-shed limit at $z=0.22$.} 
\end{figure}

At larger central redshifts, the bifurcation process changes its 
character; the discs and rings become disconnected. This is illustrated
in Fig.~10, where the bifurcation regime in the $K-\Omega$ diagram
is shown for three different values of $z$. 
At a critical value of $z$, which is
given to the numerical accuracy by $z\subscr{c} \approx 0.01$,
the connection between disc and ring is lost. Approaching
$z\subscr{c}$ from below, the curves in the $K-\Omega$ diagram begin
to be double valued until, upon reaching the critical point, they disconnect. 
Above $z\subscr{c}$, the disc sequences do not join smoothly to the
ring sequence but terminate instead at the mass-shedding limit. The
ring sequences can still be continued somewhat into the dumb-bell
regime but there the curve also terminates.
Note that the apparent difference in the bifurcation curve
for the weak relativistic limit and the Newtonian limit
(see Kley 1996, Fig.~1) stems from the fact that here
curves of constant central redshift (fixed central potential $\nu(0,0)$)
were considered, while in the Newtonian work
the curves are plotted for constant disc mass.
The Newtonian line separates the two different regimes. Curves above
end in the mass-shed limit, while curves below bifurcate continuously into
a ring sequence. 
In Fig.~11 it is demonstrated that the disc sequences for $z > z\subscr{c}$,
which terminate at the mass-shedding limit, start at that point where the
rotational velocity $v_\varphi$ at the outer rim ($\rho=\rho\subscr{d}$)
coincides with
the value of a pressure-less dust disc having the same central
redshift. Above $z\subscr{c}$, the rim-velocity lies always below that
of the dust disc.
\epsfverbosetrue
\begin{figure}
\epsfxsize=8.5cm
\epsfbox{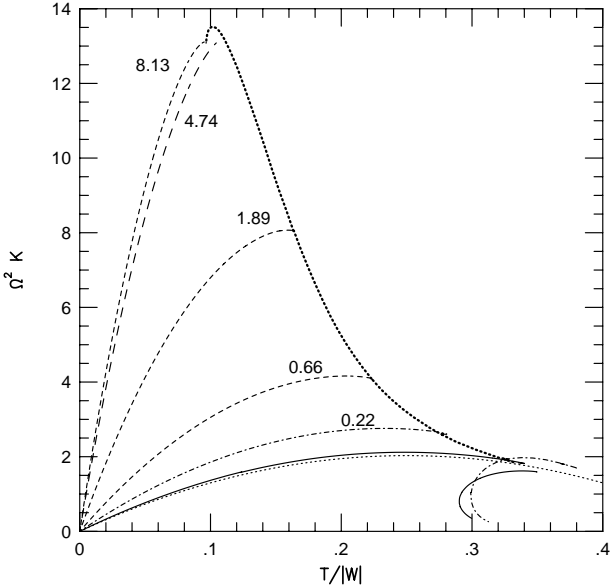}
  \caption{The normalized angular velocity versus the ratio of kinetic
  to potential energy for discs with different redshifts. The line types
  and the labeling corresponds to Fig.~7.}
\end{figure}

The total kinetic energy of the disc can be defined for rigid
rotation by
\begin{equation}
     T = \frac{1}{2} \Omega J.
\end{equation}
The total proper mass of the system, i.e. the energy of the configuration
excluding gravitational potential energy and kinetic energy, can for the
disc be defined as
\begin{equation}
     M\subscr{p} = \int\subscr{disc} ( \sigma^0 + \sigma\superscr{i} )
       u^t \sqrt{-g} \, {\rm d}^3x.
\end{equation}
The gravitational binding energy is then given by
\begin{equation}
     W = M\subscr{p} + T - M.
\end{equation} 
In Fig.~12 the normalized angular velocity is plotted versus the
ratio $t$ of the kinetic energy to the gravitational binding energy,
$t=T/|W|$.
The Newtonian curve (thin dotted line) is given by the relation
\begin{equation}
    \Omega^2 K = \frac{\pi^4}{6} t ( 1 - 2 t),
\end{equation}
which terminates at $t=0.5$. For stronger relativistic discs the
maximum possible $t$ becomes smaller and smaller because the discs reach
the mass-shed limit.
For the weaker relativistic disc the curve for the mass-shed limit
approaches the Newtonian limit tangentially.

\section{Conclusion}
We calculated the structure of uniformly rotating, infinitesimally thin,
relativistic self-gravitating discs with internal pressure. The
pressure is given by a polytropic equation of state. The polytropic
exponent $\gamma=3$ was used, since in this case 
there exists an exact solution in the Newtonian limit.
This special choice corresponds to three-dimensional bodies of
constant densities as well (see Hunter 1972). Thus, we may make
a comparison of the results found here for flat discs with those of rotating,
homogeneous relativistic stars. As Butterworth \& Ipser (1976) showed,
sequences of homogeneous, rotating relativistic stars usually
terminate at the mass-shed
limit. Close to the Newtonian case however, they were not able to follow
the sequences to this limit. We suggest that, as
weaker relativistic discs bifurcate into a ring, weaker relativistic stars
of constant density bifurcate into a toroid structure as well, and do not
end at the mass-shed limit. In the Newtonian limit, the structure of
these constant density toroids were calculated by Eriguchi \& Sugimoto (1981).

The pressure-less discs possess ergo-regions,
where the dragging of inertial frames would force observers to rotate. These
first appear at a single point within the disc at $z\subscr{e}=1.41$
and reach the edge of the disc at $z=1.89$ (Meinel \& Kleinw\"achter 1993).
They also occur in rotating stars (Butterworth \& Ipser 1976).
For discs with internal pressure however, we found no indication of
the existence of ergo-regions. It is to be expected that there exists
a continuous transition from the zero-pressure $K=0$ case to discs with
non-vanishing pressure ($K$ small). However, this connection could not be
demonstrated numerically, as disc sequences end either at the mass-shed limit
or bifurcate into rings before reaching the pressure-less limit.
One might speculate, perhaps, that
such a connection can be achieved by a disc which consists of two parts:
a pressure supported central region ($0 < \rho < \rho\subscr{p}$),
surrounded by a dust disc ($\rho\subscr{p} < \rho < \rho\subscr{d}$).
The dust disc would refer to $\rho\subscr{p}=0$, and the disc sequences
to $\rho\subscr{p}=\rho\subscr{d}$.

In some rotating relativistic stars, there are sequences of supra-massive
stars (Cook, Shapiro \& Teukolsky 1992), which are so massive that they
exceed the rest mass of a non-rotating star, and can
only exist for non-zero rotation. In the case of flat relativistic
discs with non-zero internal pressure, we did not find any
supra-massive disc sequences.
Possibly, these negative results are a result of the particular
equation of state where in the strong relativistic limit the
mass is always concentrated in the centre of the disc, as shown
in Fig.~4 for the non-rotating case.
Other relations between pressure and surface density, in particular
a smaller $\gamma$, could also possibly lead to
supra-massive sequences or ergo-regions. 

The numerical method developed here is sufficiently general to
be applied to the study of differentially
rotating discs, and can easily be extended to three-dimensional
rotating stars as well.
This may be the subject of a future further investigation. A stability
analysis of the computed equilibrium configurations lies beyond the
scope of the present paper. 
 
\section*{Acknowledgments}
Useful discussions with Drs R. Meinel and T. Wolf
are gratefully acknowledged.

\end{document}